\documentstyle[12pt]{article}
\begin{document}
%\jl{8}
%\vfill
\newtheorem{lemma}{Lemma}
\newtheorem{thm}{Theorem}
\newtheorem{cor}{Corollary}
\newtheorem{defn}{Definition}
\centerline{\Large{\bf Fractional differentiability of nowhere differentiable}} 
\centerline{\Large{\bf functions and dimensions }}
\vspace{10pt}

\centerline{ Kiran M. Kolwankar\footnote{email: kirkol@physics.unipune.ernet.in} 
and Anil D. Gangal\footnote{email: adg@physics.unipune.ernet.in}}
%\footnote{email: kirkol@physics.unipune.ernet.in} 
%\footnote{email: adg@physics.unipune.ernet.in}
\vspace{3pt}
\centerline
{\it Department of Physics, University of Pune, Pune 411 007, India.}
%\noindent
%Short title: {\bf{Fractional differentiability and dimensions}}
\vspace{6pt}
{\renewcommand{\baselinestretch}{1.95}
\begin{abstract}
Weierstrass's  everywhere continuous but
nowhere differentiable function is  shown to be locally continuously
fractionally differentiable everywhere for all orders below 
the `critical order' $2-s$ and not so for
orders between $2-s$ and $1$, where $s$, $1<s<2$ is the box dimension of the 
graph of the function.
This observation is consolidated in the
general result showing a direct connection between  local fractional
differentiability and the box dimension/ local H\"older exponent.
L\'evy index for one dimensional L\'evy flights is shown to be the
critical order of its characteristic function.
Local fractional derivatives of multifractal signals (non-random functions)
are shown to provide  the local H\"older exponent.
It is argued that Local fractional derivatives provide a powerful
tool to analyze pointwise behavior of irregular signals.
\end{abstract}
}
%\vspace{4pt}

%\noindent
%AMS Classification Scheme Numbers: {28A80, 26A33, 26A27, 26A30}
\vfill
%\noindent
%PACS Numbers: {47.53.+n, 47.52.+j, 02.30.Bi, 05.40.+j}
%\maketitle

\rightline{chao-dyn/9609016}

%\renewcommand{\baselinestretch}{2}
%\vfill
%\vspace{0.25in}
\newpage
{\bf Attractors of some dynamical systems are examples of the
occurrence of continuous but highly irregular (nondifferentiable) curves
and surfaces. Frequently their graphs are fractal sets. Ordinary calculus
is inadequate to characterize and handle such curves and surfaces. In
this paper we evolve the notion of "local fractional derivative" by
suitably modifying the concepts from fractional calculus, a branch 
which allows one to deal with derivatives and integrals of fractional
order. In particular we establish a direct quantitative connection
between the local scaling behaviour (or dimension) and the order of
differentiability. The bigger the fractal dimension the smaller is the
extent of differentiability. We show that the method developed
here provides a powerful tool for analysis of irregular and chaotic signals.
It is further noted to be suitable for dealing with fractal processes.
We have also established a local
fractional Taylor expansion, which should be of value in the approximation
of scaling signals and functions.
}
%\newpage
\section{Introduction}

The  importance of studying continuous but nowhere differentiable 
functions was emphasized a long time ago by Perrin, 
Poincar\'e and others (see Refs. \cite{1} and \cite{2}).
It is possible for a continuous function to be sufficiently irregular 
so that its graph is a fractal. This observation points out to a 
connection between the lack of differentiability of such a function and 
the dimension of its graph.
Quantitatively one would like to convert the question concerning the lack of
differentiability into one concerning the amount of loss 
of differentiability. In other words, one would 
look at derivatives of fractional order rather than only those of
integral order and relate them to dimensions.
Indeed some recent papers~\cite{3,4,5,6} indicate a connection 
between fractional calculus~\cite{7,8,9}
and fractal structure \cite{2,10} or fractal processes \cite{11,22,23}.
 Mandelbrot and Van Ness \cite{11} have 
used fractional integrals to formulate fractal processes such as fractional
Brownian motion. In Refs. \cite{4} and \cite{5} 
a fractional diffusion equation has been
proposed for the diffusion on fractals.
Also Gl\"ockle and Nonnenmacher \cite{22} have formulated fractional
differential equations for some relaxation processes which are essentially
fractal time \cite{23} processes.
Recently Zaslavasky~\cite{46} showed that the Hamiltonian chaotic dynamics
of particles can be described by a fractional generalization of the
Fokker-Plank-Kolmogorov equation which is defined by two fractional
critical exponents $(\alpha , \beta)$ responsible for the space and
time derivatives of the distribution function correspondingly.
However, to our knowledge, the precise nature of the 
connection between the dimension of the graph of a fractal curve 
and  fractional differentiability
properties has not been established.

Irregular  functions arise naturally in various 
branches of physics. It is 
well known that the graphs of projections of Brownian paths 
are nowhere differentiable and have
 dimension $3/2$. A generalization of Brownian motion called fractional 
Brownian motion \cite{2,10} gives rise to graphs having dimension between 1 and 2.
Typical Feynmann paths \cite{30,31}, like the Brownian paths are continuous but nowhere 
differentiable.
 Also, passive scalars advected by a turbulent fluid 
\cite{19,20} can have isoscalar surfaces which are highly irregular, in the limit
of the diffusion constant going to zero. Attractors of some dynamical systems
have been shown \cite{15} to be continuous but nowhere differentiable.

All these irregular functions are characterized at every point by a local H\"older
exponent typically lying between 0 and 1. 
In the case of functions having the same H\"older exponent $h$ at every 
point it is well known that the 
box dimension of its graph is $2-h$. Not all functions have the same exponent $h$
at every point but have a range of H\"older exponents.
A set $\{x\vert h(x)=h \}$ may be a fractal set.
In such situations the corresponding functions are multifractal. These kind of
functions also arise in various physical situations, for instance, velocity
field of a turbulent fluid \cite{26} at low viscosity.

 Also there exists a
class of problems where one has to solve a partial
differentiable equation subject to fractal boundary
conditions, e.g. the Laplace equation near a fractal conducting surface. 
As noted in reference \cite{39} irregular boundaries may appear, down to
a certain spatial resolution, to be non-differentiable everywhere and/or
may exhibit convolutions over many length scales.
Keeping in view such
problems there is a need to characterize pointwise behavior using something 
which can be readily used.

We consider the Weierstrass function as a prototype example of a function which is
continuous everywhere but differentiable nowhere and has an exponent which is
constant everywhere.
 One form of the Weierstrass
function is
\begin{eqnarray}
W_{\lambda}(t) = \sum_{k=1}^{\infty} {\lambda}^{(s-2)k} \sin{\lambda}^kt,\;\;\;\;\;
t \;\;\;{\rm real}.
\end{eqnarray}
For this form, when $\lambda > 1$ it is well known \cite{12} that 
$W_{\lambda}(t)$ is nowhere differentiable if $1<s<2$. 
This curve has been extensively studied \cite{1,10,13,14} and 
its graph is known to have a box dimension $s$, for sufficiently large $\lambda$.
Incidently, the Weierstrass functions are not just mathematical curiosities 
but occur at several places. For instance,
 the graph of this function is known \cite{10,15} to be a repeller
or attractor of some dynamical
systems.
This kind of function can also be recognized as 
the characteristic function of a L\'evy
flight on a one dimensional lattice \cite{16}, which means that such a L\'evy 
flight can be considered as a superposition of Weierstrass type functions.
This function has also been used \cite{10} to generate a fractional Brownian signal by multiplying 
every term by a random amplitude and randomizing phases of every term.

The main aim of the present paper is to explore the precise nature of the
connection between fractional differentiability properties of irregular 
(non-differentiable) curves and dimensions/ H\"older exponents of
their graphs. A second aim is to provide a possible tool to study
pointwise behavior.
The organization of the paper is as follows.
In section II we motivate and define what we call 
local fractional differentiability, 
formally and use a local fractional derivative to formulate the Taylor series. 
Then in section III we 
apply this definition to a specific example, viz., Weierstrass' nowhere 
differentiable function and show that this function, at every point, is
locally fractionally differentiable for  all orders below $2-s$
 and it is not so for orders between $2-s$ and 1, where $s$, $1<s<2$ 
is the box dimension of the graph of the function. In section IV we prove a
general result showing the relation between local fractional differentiability
of nowhere differentiable functions and the local H\"older exponent/
the dimension of its graph. In section V we demonstrate the use of the local
fractional derivatives (LFD) in unmasking isolated singularities and
in the study of the pointwise behavior of multifractal functions.
In section VI  we conclude after pointing out a few possible consequences of
our results.

\section{Fractional Differentiability}

 We begin by recalling the Riemann-Liouville definition of the 
fractional integral
of a real function, which is
given by \cite{7,9}
\begin{eqnarray}
{{d^qf(x)}\over{[d(x-a)]^q}}={1\over\Gamma(-q)}{\int_a^x{{f(y)}\over{(x-y)^{q+1}}}}dy
\;\;\;{\rm for}\;\;\;q<0,\;\;\;a\;\;{\rm real},\label{def1}
\end{eqnarray}
and of the fractional derivative
\begin{eqnarray}
{{d^qf(x)}\over{[d(x-a)]^q}}={1\over\Gamma(1-q)}{d\over{dx}}{\int_a^x{{f(y)}\over{(x-y)^{q}}}}dy
\;\;\;{\rm for}\;\;\; 0<q<1.\label{def2}
\end{eqnarray}
The case of $q>1$ is of no relevance in this paper.
For future reference we note \cite{7,9}
\begin{eqnarray}
{{d^qx^p}\over {d x^q}} = {\Gamma(p+1) \over {\Gamma(p-q+1)}} x^{p-q}\;\;\;
{\rm for}\;\;\;p>-1.\label{xp}
\end{eqnarray}
We also note that the fractional derivative has the property (see Ref.
 \cite{7}), viz.,
\begin{eqnarray}
{d^qf(\beta x)\over{d x^q}}={{\beta}^q{d^qf(\beta x)\over{d(\beta x)^q}}}
\end{eqnarray}
which makes it suitable for the study of scaling.

One may note that except in the case of positive integral $q$, the $q$th derivative
will be nonlocal through its dependence
on the lower limit "$a$".
 On the other hand we wish to
study local scaling properties and hence we need to introduce the notion
of local fractional differentiability.
Secondly from Eq. (\ref{xp}) it is clear that the fractional derivative of a constant
function is not zero.
These two features play an
important role in defining local fractional differentiability. 
We note that changing the lower
limit or adding a constant to a function alters the value of the fractional
derivative. This forces one to choose the lower limit as well as the additive
constant before hand. The most natural choices are as follows.
(1) We subtract, from the function, the value of the function at the point where
 fractional differentiability is to be checked. This makes the value of the function
zero at that point, washing out the effect of any constant term.
 (2) The natural choice of a lower limit will be
that point, where we intend to examine the fractional differentiability, itself.
 This has an advantage in that it preserves local nature of
the differentiability property. With these motivations we now introduce
the following.

\begin{defn} If, for a function $f:[0,1]\rightarrow I\!\!R$, the  limit 
\begin{eqnarray}
I\!\!D^qf(y) = 
{\lim_{x\rightarrow y} {{d^q(f(x)-f(y))}\over{d(x-y)^q}}},\label{defloc}
\end{eqnarray}
exists and is finite, then we say that the {\it local fractional derivative} (LFD) 
of order $q$, at $x=y$, 
exists. 
\end{defn}
\begin{defn}
We define {\it critical order} $\alpha$, at $y$, as
$$
\alpha(y) = Sup \{q \vert {\rm {all\;local\; fractional\; derivatives\; of\; order\; less\; than\;}} q{{\rm\; exist\; at}\;y}\}. 
$$
\end{defn}
Incidentally we note that Hilfer \cite{17,18} used a similar notion to extend 
Ehrenfest's classification of phase
transition to continuous transitions. However in his work only the singular part of 
the free energy was considered. So the first of the above mentioned 
condition was automatically 
satisfied. Also no lower limit of fractional derivative was considered 
and by default it was taken as zero.

In order to see the information contained in the LFD we consider the
fractional Taylor's series with a remainder term for a real function $f$.
Let
\begin{eqnarray}
F(y,x-y;q) = {d^q(f(x)-f(y))\over{[d(x-y)]^q}}.
\end{eqnarray}
It is clear that
\begin{eqnarray}
I\!\!D^qf(y)=F(y,0;q).
\end{eqnarray}
Now, for $0<q<1$,
\begin{eqnarray}
f(x)-f(y)& =& {1\over\Gamma(q)} \int_0^{x-y} {F(y,t;q)\over{(x-y-t)^{-q+1}}}dt\\
&=& {1\over\Gamma(q)}[F(y,t;q) \int (x-y-t)^{q-1} dt]_0^{x-y} \nonumber\\
&&\;\;\;\;\;\;\;\;+ {1\over\Gamma(q)}\int_0^{x-y} {dF(y,t;q)\over{dt}}{(x-y-t)^q\over{q}}dt,
\end{eqnarray}
provided the last term exists. Thus
\begin{eqnarray}
f(x)-f(y)&=& {I\!\!D^qf(y)\over \Gamma(q+1)} (x-y)^q \nonumber\\
&&\;\;\;\;\;\;\;\;+ {1\over\Gamma(q+1)}\int_0^{x-y} {dF(y,t;q)\over{dt}}{(x-y-t)^q}dt,\label{taylor}
\end{eqnarray}
i.e.
\begin{eqnarray}
f(x) = f(y) + {I\!\!D^qf(y)\over \Gamma(q+1)} (x-y)^q + R_1(x,y),\label{taylor2}
\end{eqnarray}
where $R_1(x,y)$ is a remainder given by
\begin{eqnarray}
R_1(x,y) = {1\over\Gamma(q+1)}\int_0^{x-y} {dF(y,t;q)\over{dt}}{(x-y-t)^q}dt
\end{eqnarray}
Equation (\ref{taylor2}) is a fractional Taylor expansion of $f(x)$ involving
only the lowest and the second leading terms. This expansion can be carried
to higher orders provided the corresponding remainder term is well defined.

We note that the local fractional derivative as defined above
(not just the fractional derivative) provides
the coefficient $A$ in the approximation
of $f(x)$ by the function $f(y) + A(x-y)^q/\Gamma(q+1)$, for $0<q<1$, 
in the vicinity of $y$. 
 We further note that the terms
on the RHS of Eq. (\ref{taylor}) are non-trivial and finite only in the case 
$q=\alpha$.
Osler in Ref.\cite{21} has constructed a fractional Taylor
series using usual (not local in the present sense) fractional derivatives. 
His results are, however, applicable to analytic functions and cannot be 
used for non-differentiable scaling functions directly. Further Osler's
formulation involves terms with negative $q$ also and hence is not suitable
for approximating schemes.

 One may further notice that when $q$ is set equal to
one in the above approximation one gets
the equation of the tangent. 
It may be recalled that all the curves passing through a point $y$ and having 
the same tangent
form an equivalence class (which is modeled by a linear behavior). 
Analogously all the functions (curves) with the same critical order $\alpha$
and the same $I\!\!D^{\alpha}$
will form an equivalence class modeled by $x^{\alpha}$ [ 
If $f$ differs from $x^{\alpha}$ by a logarithmic correction then 
terms on RHS of Eq. (\ref{taylor})
do not make sense precisely as in the case of ordinary calculus]. 
This is how one may
generalize the geometric interpretation of derivatives in terms of tangents.  
This observation is useful when one wants to approximate an irregular 
function by a piecewise smooth (scaling) function.

To illustrate the definitions of local fractional differentiability and critical order 
consider an example of a polynomial of degree $n$ with its graph passing through the 
origin and for which the first derivative at the origin
is not zero. Then all the local fractional derivatives of order
less than or equal to one exist at the origin. Also all derivatives 
of integer order greater than
one exist, as expected. But local derivatives of any other order,
 e.g. between 1 and 2 [see
equations (\ref{xp}) and (\ref{defloc})] do not exist. 
Therefore critical order for this function at
$x=0$ is one. In fact, except at a finite number of points 
where the function has a
vanishing first derivative, critical order
of a polynomial function will be one, since the linear term is expected to
dominate near these points.

\noindent
{\bf Remark}: We would like to point out that 
there is a multiplicity of definitions of a fractional derivative.
The use of a Riemann-Liouville
 definition, and other equivalent definitions such as Grunwald's  
definition, are suitable for our purpose.
 The other definitions of fractional derivatives which
do not allow control over both the limits, such as Wyel's definition or definition
using Fourier transforms, are not suitable since
it would not be possible to retrieve the local nature of
the differentiability property which is essential for the study of
local behavior. Also, the important difference between our work and
the work of \cite{4,22} is that while we are trying to study the local scaling behavior these works apply to asymptotic scaling properties.

\section{Fractional Differentiability of Weierstrass Function}

Consider a form of the Weierstrass function as given above, viz.,
\begin{eqnarray}
W_{\lambda}(t) = \sum_{k=1}^{\infty} {\lambda}^{(s-2)k} 
\sin{\lambda}^kt,\;\;\;\;
\lambda>1.
\end{eqnarray}
Note that $W_{\lambda}(0)=0$. 
Now
\begin{eqnarray}
{{d^qW_{\lambda}(t)}\over{dt^q}}  
&=& {\sum_{k=1}^{\infty} {\lambda}^{(s-2)k}{{d^q\sin({\lambda}^kt)}\over {dt^q}}}\nonumber\\
&=& {\sum_{k=1}^{\infty} {\lambda}^{(s-2+q)k}{{d^q\sin({\lambda}^kt)}\over {d({\lambda}^kt)^q}}}, \nonumber
\end{eqnarray}
provided the right hand side converges uniformly. Using, for $0<q<1$, 
\begin{eqnarray}
{{d^q\sin(x)}\over {d x^q}}={{d^{q-1}\cos(x)}\over{d x^{q-1}}}, \nonumber
\end{eqnarray}
 we get
\begin{eqnarray}
{{d^qW_{\lambda}(t)}\over{dt^q}}  
&=& {\sum_{k=1}^{\infty} {\lambda}^{(s-2+q)k}{{d^{q-1}cos({\lambda}^kt)}\over {d({\lambda}^kt)^{q-1}}}}\label{a}
\end{eqnarray}
From the second mean value theorem it follows that the fractional integral 
of $\cos({\lambda}^kt)$ of order $q-1$ is 
bounded uniformly for all values of ${\lambda}^kt$. 
This implies that the series on the right 
hand side will converge uniformly for $q<2-s$, justifying our action of taking
the fractional derivative operator inside the sum.

Also as $t \rightarrow 0$ for
any $k$ the fractional integral in the summation of equation (\ref{a}) goes to zero.
Therefore it is easy to see from this that
\begin{eqnarray}
I\!\!D^qW_{\lambda}(0) = {\lim_{t\rightarrow 0} {{d^qW_{\lambda}(t)}\over{dt^q}}}=0\;\;\;
{\rm for} \;\;\;q<2-s.
\end{eqnarray}
This shows that the $q${th} local derivative of the Weierstrass function exists and
is continuous, at $t=0$, for $q<2-s$.

To check the fractional differentiability at any other point, say $\tau$,
we use $t'=t-\tau$ and $\widetilde{W} (t' )=W(t'+\tau )-W(\tau)$ so that
$\widetilde{W}(0)=0$. We have
\begin{eqnarray}
\widetilde{W}_{\lambda} (t' ) &=& \sum_{k=1}^{\infty} {\lambda}^{(s-2)k} \sin{\lambda}^k(t' +\tau)-
\sum_{k=1}^{\infty} {\lambda}^{(s-2)k} \sin{\lambda}^k\tau \nonumber\\
&=&\sum_{k=1}^{\infty} {\lambda}^{(s-2)k}(\cos{\lambda}^k\tau \sin{\lambda}^kt' +
\sin{\lambda}^k\tau(\cos{\lambda}^kt' -1)). \label{c}
\end{eqnarray}
Taking the fractional derivative of this with respect to $t'$ and following the 
same procedure we can show that the fractional derivative of the Weierstrass
function of order $q<2-s$ exists at all points.

For $q>2-s$,  right hand side of the equation (\ref{a})  seems to diverge.
We now prove that the LFD of order $q>2-s$ in fact does not
exist.
We do this by showing that there exists a sequence of 
points approaching 0 along which
the limit of the fractional derivative of order $2-s < q <1$ does not exist.
We use the property of the Weierstrass function \cite{10}, viz.,
for each $t' \in [0,1]$ and $0 < \delta \leq {\delta}_0$
there exists $t$ such that $\vert t-t' \vert \leq \delta$ and
\begin{eqnarray}
c{\delta}^{\alpha} \leq \vert W(t)-W(t') \vert , \label{uholder}
\end{eqnarray}
where $c > 0$ and $\alpha=2-s$, provided $\lambda$ is sufficiently large.
 We consider the case of $t'=0$ and
$t>0$.
Define $g(t)=W(t)-ct^{\alpha}$.

Now the above mentioned property, along with  continuity
of the Weierstrass function assures us a 
sequence of points $t_1>t_2>...>t_n>...\geq 0$ such that
$t_n \rightarrow 0$ as $n \rightarrow \infty$ and $g(t_n) = 0$ 
and $g(t)>0$ on $(t_n,\epsilon)$ for some $\epsilon>0$, for all
$n$ (it is not ruled out that $t_n$ may be zero for finite $n$). 
Define
\begin{eqnarray}
g_n(t)&=&0,\;\;\;{\rm if}\;\;\;t\leq t_n, \nonumber\\
&=&g(t),\;\;\; {\rm otherwise}.\nonumber
\end{eqnarray}
Now we have, for $0 <\alpha < q < 1$,
\begin{eqnarray}
{{d^qg_n(t)}\over{d(t-t_n)^q}}={1\over\Gamma(1-q)}{d\over{dt}}{\int_{t_n}^t{{g(y)}\over{(t-y)^{q}}}}dy,\nonumber
\end{eqnarray}
where $t_n \leq t \leq t_{n-1}$. We assume that the left hand side of the above
equation exists for if it does not then we have nothing to prove.
Let
\begin{eqnarray}
h(t)={\int_{t_n}^t{{g(y)}\over{(t-y)^{q}}}}dy.\nonumber
\end{eqnarray}
Now $h(t_n)=0$ and $h(t_n+\epsilon)>0$, for a suitable $\epsilon$, as the integrand is positive.
Due to continuity there must exist an ${\epsilon}'>0$ and ${\epsilon}'<\epsilon $
such that $h(t)$ is increasing on $(t_n,{\epsilon}')$.
Therefore
\begin{eqnarray}
0 \leq {{d^qg_n(t)}\over{d(t-t_n)^q}} {\vert}_{t=t_n},\;\;\;\;n=1,2,3,...  .
\end{eqnarray}
This implies that
\begin{eqnarray}
c{{d^qt^{\alpha}}\over{d(t-t_n)^q}} {\vert}_{t=t_n} \leq {{d^qW(t)}\over{d(t-t_n)^q}} {\vert}_{t=t_n},
\;\;\;\;n=1,2,3,...  .
\end{eqnarray}
But we know from Eq. (\ref{xp}) that, when $0<\alpha <q<1$,
the left hand side in the above inequality approaches infinity as $t\rightarrow 0$.
This implies that the right hand side of the above inequality does not
exist as $t \rightarrow 0$. This argument can be generalized 
for all non-zero $t'$ by
changing the variable $t''=t-t'$.
This concludes the proof.

Therefore the critical order of the Weierstrass function 
will be $2-s$ at all points.

\noindent
{\bf Remark}: Schlesinger et al \cite{16} have considered a 
L\'evy flight on a one dimensional 
periodic lattice where a particle jumps from one lattice site 
to other  with the probability given by
\begin{eqnarray}
P(x) = {{{\omega}-1}\over{2\omega}} \sum_{j=0}^{\infty}{\omega}^{-j}
[\delta(x, +b^j) + \delta(x, -b^j)],
\end{eqnarray}
where $x$ is magnitude of the jump, $b$ is a lattice spacing and $b>\omega>1$. 
$\delta(x,y)$ is a Kronecker delta.
The characteristic function for $P(x)$ is given by
\begin{eqnarray}
\tilde{P}(k) = {{{\omega}-1}\over{2\omega}} \sum_{j=0}^{\infty}{\omega}^{-j}
\cos(b^jk).
\end{eqnarray}
which is nothing but the Weierstrass cosine function.
For this distribution the L\'evy index is $\log{\omega}/\log{b}$, which can be
identified as the critical order of $\tilde{P}(k)$. 

More generally for the L\'evy distribution with index $\mu$ 
the characteristic function
is given by
\begin{eqnarray}
\tilde{P}(k) =A \exp{c\vert k \vert^{\mu}}.
\end{eqnarray}
The critical order of this function at $k=0$
also turns out to be same as $\mu$. Thus the L\'evy index can be identified as
the critical order of the characteristic function at $k=0$.

\section{Connection between critical order and the box dimension of the curve}

\begin{thm}
 Let $f:[0,1]\rightarrow I\!\!R$ be a continuous function.

a) If
\begin{eqnarray}
\lim_{x\rightarrow y} {d^q(f(x)-f(y)) \over{[d(x-y)]^q}}=0,\;\;\;
{\rm for}\;\; q<\alpha\;\; 
,\nonumber
\end{eqnarray}
where $q,\alpha \in (0,1)$,
for all $y \in (0,1)$, 
then  $dim_Bf(x) \leq 2-\alpha$.

b) If there exists a sequence  $x_n \rightarrow y$ as
$n \rightarrow \infty$ such that
\begin{eqnarray}
\lim_{n\rightarrow \infty} {d^q(f(x_n)-f(y)) \over{[d(x_n-y)]^q}}=\pm \infty,\;\;\;
{\rm for}\;\; q>\alpha,\;\; 
,\nonumber
\end{eqnarray}
for all $y$, 
then $dim_Bf \geq 2-\alpha$.
\end{thm}

\noindent
{\bf Proof}: (a) Without loss of generality assume $y=0$ and $f(0)=0$.
We consider the case of $q<\alpha$.

As $0<q<1$ and $f(0)=0$ we can write \cite{7} 
\begin{eqnarray}
f(x)&=&{d^{-q}\over{d x^{-q}}}{d^qf(x)\over{d x^q}}\nonumber\\ 
&=&{1\over\Gamma(q)}{\int_0^x{{d^qf(y)\over{dy^q}}\over{(x-y)^{-q+1}}}}dy. \label{comp}
\end{eqnarray}
Now
\begin{eqnarray}
\vert f(x)\vert \leq {1\over\Gamma(q)}{\int_0^x{\vert {d^qf(y)\over{dy^q}}\vert
\over{(x-y)^{-q+1}}}}dy. \nonumber
\end{eqnarray}
As, by assumption, for $q<\alpha$,
\begin{eqnarray}
\lim_{x\rightarrow 0}{d^qf(x)\over{d x^q}}=0,\nonumber
\end{eqnarray}
 we have, for any $\epsilon > 0$, a $\delta > 0$ such that
$\vert {d^qf(x)/{d x^q}}\vert < \epsilon$ for all $x< \delta$,
\begin{eqnarray}
\vert f(x)\vert &\leq& {\epsilon\over\Gamma(q)}{\int_0^x{dy
\over{(x-y)^{-q+1}}}}\nonumber\\
&=&{\epsilon\over \Gamma(q+1)}x^q.\nonumber
\end{eqnarray}
As a result we have
\begin{eqnarray}
\vert f(x)\vert &\leq& K \vert x\vert ^q, \;\;\;\;{\rm for}\;\;\; x<\delta.\nonumber
\end{eqnarray}
Now this argument can be extended for general $y$ simply by considering
$x-y$ instead of $x$ and $f(x)-f(y)$ instead of $f(x)$. So finally
we get for $q<\alpha$
\begin{eqnarray}
\vert f(x)-f(y)\vert &\leq& K \vert x-y\vert ^q, \;\;\;\;{\rm for}
\;\;\vert x-y \vert < \delta,\label{holder}
\end{eqnarray}
for all $y \in (0,1)$. Hence we have \cite{10}
\begin{eqnarray}
{\rm dim}_Bf(x) \leq 2-\alpha.\nonumber
\end{eqnarray}

b) Now we consider the case $q>\alpha$. If we have
\begin{equation}
\lim_{x_n\rightarrow 0}{d^qf(x_n)\over{dx_n^q}}=\infty, \label{k0}
\end{equation}
then for given $M_1 >0$ and $\delta > 0$ we can find positive integer $N$ such that $|x_n|<\delta$ and
$ {d^qf(x_n)}/{dx_n^q} \geq M_1$ for all $n>N$. Therefore by Eq. (\ref{comp})
\begin{eqnarray}
 f(x_n) &\geq& {M_1\over\Gamma(q)}{\int_0^{x_n}{dy
\over{(x_n-y)^{-q+1}}}}\nonumber\\
&=&{M_1\over \Gamma(q+1)}x_n^q\nonumber
\end{eqnarray}
If we choose $\delta=x_N$ then we can say that there exists $x<\delta$
such that
\begin{eqnarray}
f(x) \geq k_1 {\delta}^q. \label{k1}
\end{eqnarray}
If we have
\begin{eqnarray}
\lim_{x_n\rightarrow 0}{d^qf(x_n)\over{dx_n^q}}=-\infty, \nonumber
\end{eqnarray}
then for given $M_2 >0$ we can find a positive integer $N$ such that
$ {d^qf(x_n)}/{dx_n^q} \leq -M_2$ for all $n>N$. Therefore
\begin{eqnarray}
 f(x_n) &\leq& {-M_2\over\Gamma(q)}{\int_0^{x_n}{dy
\over{(x_n-y)^{-q+1}}}}\nonumber\\
&=&{-M_2\over \Gamma(q+1)}x_n^q.\nonumber
\end{eqnarray}
Again if we write $\delta=x_N$, there exists $x<\delta$ such that
\begin{eqnarray}
f(x) \leq  -k_2 {\delta}^q.\label{k2}
\end{eqnarray}
Therefore by (\ref{k1}) and (\ref{k2}) there exists $x<\delta$ such that, for $q>\alpha$, 
\begin{eqnarray}
\vert f(x)\vert &\geq& K \delta^q.\nonumber
\end{eqnarray}
Again for any $y \in (0,1)$ there exists $x$ such that
for $q>\alpha$ and $|x-y|<\delta$
\begin{eqnarray}
\vert f(x)-f(y)\vert &\geq& k \delta^q.\nonumber
\end{eqnarray}
 Hence we have \cite{10}
\begin{eqnarray}
{\rm dim}_Bf(x) \geq 2-\alpha.\nonumber
\end{eqnarray}

Notice that  part (a) of the theorem above is the generalization
of the statement that $C^1$ functions are locally Lipschitz (hence their
graphs have dimension 1) to the case when the function has a H\"older type
upper bound (hence their dimension is greater than one).

Here the function is required to
have the same critical order throughout the interval. We can weaken this
condition slightly. Since we are dealing with a box dimension which
is  finitely stable \cite{10}, we can allow a finite number of points having
different critical order so that we can divide the set in finite parts
having the same critical order in each part.

The example of a polynomial of degree $n$ having critical order one and dimension one is 
consistent with the above result, as we can divide the graph of the polynomial 
in a finite
number of parts such that at each point in every part the critical order is one.
Using the finite stability of the box dimension, the dimension of the whole curve
will be one.

We can also prove a partial converse of the above theorem.

\begin{thm} 

Let $f:[0,1]\rightarrow I\!\!R$ be a continuous function.

a) Suppose 
\begin{eqnarray}
\vert f(x)- f(y) \vert \leq c\vert x-y \vert ^{\alpha}, \nonumber
\end{eqnarray}
where $c>0$, $0<\alpha <1$ and $|x-y|< \delta$ for some $\delta >0$.
Then
\begin{eqnarray}
\lim_{x\rightarrow y} {d^q(f(x)-f(y)) \over{[d(x-y)]^q}}=0,\;\;\;
{\rm for}\;\; q<\alpha,\;\; 
\nonumber
\end{eqnarray}
for all $y\in (0,1)$.

b) Suppose that for each $y\in (0,1)$ and for each $\delta >0$ there exists x such that
$|x-y| \leq \delta $ and
\begin{eqnarray}
\vert f(x)- f(y) \vert \geq c{\delta}^{\alpha}, \nonumber
\end{eqnarray}
where $c>0$, $\delta \leq {\delta}_0$ for some ${\delta}_0 >0$ and $0<\alpha<1$.
Then there exists a sequence $x_n \rightarrow y$ as $n\rightarrow \infty$
such that
\begin{eqnarray}
\lim_{n\rightarrow \infty} {d^q(f(x_n)-f(y)) \over{[d(x_n-y)]^q}}=\pm \infty,\;\;\;
{\rm for}\;\; q>\alpha,\;\; 
\nonumber
\end{eqnarray}
for all $y$.
\end{thm}

\noindent
{\bf Proof}

a) Assume that there exists a sequence  $x_n \rightarrow y$ as
$n \rightarrow \infty$ such that
\begin{eqnarray}
\lim_{n\rightarrow \infty} {d^q(f(x_n)-f(y)) \over{[d(x_n-y)]^q}}=\pm \infty,\;\;\;
{\rm for}\;\; q<\alpha\;\;,\nonumber
\end{eqnarray}
for some $y$. Then by arguments between Eq. (\ref{k0}) and Eq. (\ref{k1}) of the second part of the previous theorem it is a 
contradiction.
Therefore
\begin{eqnarray}
\lim_{x\rightarrow y} {d^q(f(x)-f(y)) \over{[d(x-y)]^q}}={\rm const}\;\;
{\rm or}\;\; 0,\;\;\;
{\rm for}\;\; q<\alpha.
\nonumber
\end{eqnarray}
Now if
\begin{eqnarray}
\lim_{x\rightarrow y} {d^q(f(x)-f(y)) \over{[d(x-y)]^q}}={\rm const},\;\;\;
{\rm for}\;\; q<\alpha,\;\; 
\nonumber
\end{eqnarray}
then we can write
\begin{eqnarray}
{d^q(f(x)-f(y)) \over{[d(x-y)]^q}}=K+\eta(x,y),\;\;\;
\nonumber
\end{eqnarray}
where $K={\rm const}$ and $\eta(x,y) \rightarrow 0$ 
sufficiently fast as $x\rightarrow y$. Now
taking the $\epsilon$ derivative of both sides, 
for sufficiently small $\epsilon$ we get
\begin{eqnarray}
 {d^{q+\epsilon}(f(x)-f(y)) \over{[d(x-y)]^{q+\epsilon}}}={{K(x-y)^{-\epsilon}}
\over {\Gamma(1-\epsilon)}} + {d^{\epsilon}{\eta(x,y)}\over{[d(x-y)]^{\epsilon}}}
\;\;\;{\rm for}\;\; q+\epsilon <\alpha. \nonumber
\end{eqnarray}
As $x\rightarrow y$ the right hand side of the above equation goes
to infinity (the term involving $\eta$ does not matter since $\eta$ goes to 0
sufficiently fast)
which again is a contradiction. Hence the proof.

b)The proof follows by the method used in the previous section to show that
the fractional derivative of order greater than $2-\alpha$ of the Weierstrass
function does not exist.

These two theorems give an equivalence between the H\"older exponent and the critical
order of fractional differentiability.

\section{Local Fractional Derivative as a tool to study pointwise regularity of functions}

Motivation for studying pointwise behavior of irregular functions
 and its relevance in physical
processes was given in the Introduction.
There are several approaches to
studying the pointwise behavior of functions. Recently wavelet transforms \cite{29,38}
were used for this purpose and  have met with some success. 
In this section we argue that LFDs is a tool that can be used to characterize
irregular functions and has certain advantages over its 
counterpart using wavelet transforms in aspects explained below. 
Various authors \cite{27,24} have used the following general definition
of H\"older exponent. The H\"older exponent $\alpha(y)$ of a function $f$
at $y$ is defined as the largest exponent such that there exists a polynomial
$P_n(x)$ of order $n$ that satisfies
\begin{eqnarray}
\vert f(x) - P_n(x-y) \vert = O(\vert x-y \vert^{\alpha}),
\end{eqnarray}
for $x$ in the neighborhood of $y$. This definition is equivalent to
equation (\ref{holder}), for $0<\alpha<1$, the range of interest in this work.

It is clear from theorem I that
LFDs provide an algorithm to calculate H\"older exponents and 
dimensions. It may be noted that since there is a clear change
in behavior when order $q$ of the derivative crosses the critical order
of the function 
it should be easy to determine the H\"older exponent numerically.
Previous methods using autocorrelations for fractal signals \cite{10} 
involve an additional step of finding an autocorrelation.

\subsection{Isolated singularities and masked singularities}

  Let us first consider the case of isolated
singularities. We choose the simplest example $f(x)=ax^{\alpha},\;\;\;0<\alpha
<1,\;\;\;x>0$.  The critical order at $x=0$ gives the order of 
singularity at that point whereas
the value of the LFD $I\!\!D^{q=\alpha}f(0)$, viz 
$a\Gamma(\alpha+1)$, gives the strength of the singularity.

Using LFD we can detect a weaker singularity masked by a stronger singularity.
As demonstrated below, we can estimate and subtract the contribution due to 
the stronger singularity from the 
function and find out the critical order of the remaining function.
Consider, for example, the function
\begin{eqnarray}
f(x)=ax^{\alpha}+bx^{\beta},\;\;\;\;\;\;0<\alpha <\beta <1,\;\;\;x>0.
\label{masked}
\end{eqnarray}
The LFD of this function at $x=0$ of the order $\alpha$ is 
$I\!\!D^{\alpha}f(0)=a\Gamma(\alpha+1)$.
Using this estimate of stronger singularity we now write
 $$
G(x;\alpha)=f(x)-f(0)-{I\!\!D^{\alpha}f(0)\over\Gamma(\alpha+1)}x^{\alpha},
$$
which for the function $f$ in Eq. (\ref{masked}) is
\begin{eqnarray}
{ {d^q G(x'\alpha) }
\over{d x^q}} = {b\Gamma(\beta+1)\over{\Gamma(\beta-q+1)}}x^{\beta-q}.
\end{eqnarray}
Therefore the critical order of the  function $G$, at $x=0$, is $\beta$. 
 Notice that the estimation of the weaker singularity was possible in the
above calculation just because the LFD gave the coefficient of $x^{\alpha}/
{\Gamma(\alpha+1)}$. This suggests that using LFD, one should be able to extract the secondary singularity spectrum
masked by the primary singularity spectrum of strong singularities. Hence one 
can gain more insight into the processes giving rise to irregular
behavior. Also, one may note that this procedure can be used to detect
singularities masked by regular polynomial behavior. In this way one can extend
the present analysis beyond the range $0<\alpha<1$, where $\alpha$ is a H\"older 
exponent.

A comparison of the two methods of studying pointwise behavior 
of functions, one using wavelets and the other using LFD, 
shows that characterization of H\"older classes of
functions using LFD is direct and involves fewer assumptions. 
The characterization of a H\"older class of functions with 
oscillating singularity, 
e.g. $f(x)=x^{\alpha}\sin(1/x^{\beta})$ ($x>0$, $0< \alpha <1$ and $\beta>0$), 
using wavelets needs two exponents \cite{25}. 
Using LFD, owing to theorem I and II critical order 
directly  gives the H\"older exponent for such a function.  

It has 
been shown in the context of wavelet transforms that 
one can detect singularities masked by regular polynomial
behavior \cite{27} by choosing the analyzing wavelet with 
its first $n$ (for suitable $n$)
moments vanishing. If one has to extend the wavelet method 
for the unmasking of weaker singularities,
 one would then require analyzing wavelets with fractional moments vanishing.
 Notice that
 one may require this condition along with the condition 
on the first $n$ moments. Further the class of functions to be analyzed is in
general restricted in these analyses. These restrictions essentially arise
from the asymptotic properties of the wavelets used.
 On the other hand, with the truly
local nature of LFD one does not have to bother about the behavior of functions
outside our range of interest.

\subsection{Treatment of multifractal function}

Multifractal measures have been the object of many investigations
\cite{32,33,34,35,40}. This 
formalism has met with many applications. Its importance also stems
from the fact such measures are natural measures to be used in the
analysis of many phenomenon \cite{36,37}. It may however happen that the object
one wants to understand is a function (e.g., a fractal or multifractal signal)
rather than a set or a measure. For instance one would like to
characterize the velocity of fully developed turbulence. We now proceed
with the analysis of such multifractal functions using LFD. 

 Now we consider the case 
of multifractal functions. Since LFD gives the local 
and pointwise behavior of the function, conclusions of theorem I will
carry over even in the case of multifractal functions where we have 
different H\"older exponents at different points.
Multifractal functions have been defined by Jaffard \cite{24}
 and Benzi et al. \cite{28}. 
However as noted by  Benzi et al. their functions are random in nature and 
the pointwise behavior
can not be studied. Since we are dealing with non-random 
functions in this paper,
we shall consider a specific (but non-trivial) example of a function 
constructed by Jaffard to illustrate the procedure. This function is a 
solution $F$ of the functional equation 
\begin{eqnarray} 
F(x)=\sum_{i=1}^d {\lambda}_iF(S_i^{-1}(x)) + g(x),
\end{eqnarray}
where $S_i$'s are the affine transformations of the kind 
$S_i(x)={\mu}_ix+b_i$ (with $\vert \mu_i \vert < 1$ and $b_i$'s real)
 and  $\lambda_i$'s 
are some real numbers and $g$ is any sufficiently smooth function  ($g$ and its
 derivatives should have a fast decay). For the sake of illustration 
we choose ${\mu}_1={\mu}_2=1/3$, $b_1=0$, $b_2=2/3$, 
${\lambda}_1=3^{-\alpha}$, ${\lambda}_2=3^{-\beta}$ ($0<\alpha<\beta<1$) and 
\begin{eqnarray}
g(x)&=& \sin(2\pi x),\;\;\;\;\;\;{\rm if}\;\;\;\; x\in [0,1],\nonumber\\
&=&0,\;\;\;\;\;\;\;\;\;{\rm otherwise}. \nonumber
\end{eqnarray}
Such functions are studied in detail in Ref. \cite{24} using wavelet transforms
where it has been shown that  the  above functional equation (with the
parameters we have chosen)
has a unique solution $F$ and at any point
$F$ either has H\"older exponents ranging from 
$\alpha$ to $\beta$ or is smooth.  A sequence of points $S_{i_1}(0),\;\;$ 
$\;S_{i_2}S_{i_1}(0),\;\;$ 
$\cdots,\;\;\; S_{i_n}\cdotp \cdotp \cdotp S_{i_1}(0), \;\;\cdots$,
where $i_k$ takes values 1 or 2, 
tends to a point in $[0,1]$ (in fact to a point of a triadic
cantor set) and for the values of 
${\mu}_i$s we have chosen this correspondence between sequences and limits 
is one to one. 
The solution of the above functional equation is given by Ref. \cite{24} as
\begin{eqnarray}
F(x)=\sum_{n=0}^{\infty}\;\;\;\sum_{i_1,\cdots,i_n=1}^2{\lambda}_{i_1}\cdots{\lambda}_{i_n}
g(S_{i_n}^{-1}\cdots S_{i_1}^{-1}(x)). \label{soln}
\end{eqnarray}
Note that with the above choice of parameters the inner sum in (\ref{soln})
reduces to a single term. Jaffard \cite{24} has shown that
\begin{eqnarray}
h(y)=\liminf_{n\rightarrow \infty}{{\log{({\lambda}_{{i_1}(y)}\cdots{\lambda}_{{i_n}(y)})}}
\over{\log{({\mu}_{{i_1}(y)}\cdots{\mu}_{{i_n}(y)})}}},
\end{eqnarray}
where $\{i_1(y)\cdot\cdot\cdot i_n(y)\}$ is a sequence of integers appearing in 
the sum in equation (\ref{soln}) at a point $y$, 
and is the local H\"older exponent at $y$.
 It is clear that $h_{min}=\alpha$ and
$h_{max}=\beta$. The function $F$ at the points of a triadic cantor
 set have $h(x) \in [\alpha , \beta]$
and at other points it is smooth ( where $F$ is as smooth as $g$).
Now 
\begin{eqnarray}
{d^q(F(x)-F(y))\over{[d(x-y)]^q}}&=&\sum_{n=0}^{\infty}\;\;\;\sum_{i_1,\cdots,i_n=1}^2{\lambda}_{i_1}\cdots{\lambda}_{i_n}\nonumber\\
&&\;\;\;\;\;\;\;\;\;{d^q[g(S_{i_n}^{-1}\cdots S_{i_1}^{-1}(x))-g(S_{i_n}^{-1}\cdots S_{i_1}^{-1}(y))]
\over{[d(x-y)]^q}}\nonumber\\
&=&\sum_{n=0}^{\infty}\;\;\;\sum_{i_1,\cdots,i_n=1}^2{\lambda}_{i_1}\cdots{\lambda}_{i_n}
({\mu}_{i_1}\cdots{\mu}_{i_n})^{-q} \nonumber\\
&&\;\;\;\;\;\;\;\;\;\;{d^q[g(S_{i_n}^{-1}\cdots S_{i_1}^{-1}(x))-g(S_{i_n}^{-1}\cdots S_{i_1}^{-1}(y))]
\over{[d(S_{i_n}^{-1}\cdots S_{i_1}^{-1}(x-y))]^q}}, \label{fdj}
\end{eqnarray}
provided the RHS is uniformly bounded.
Following the procedure described in section III the fractional
derivative on the RHS can easily be seen to be uniformly bounded and 
the series is convergent if $q<\min\{h(x),h(y)\}$.
Further it vanishes in the limit as $x\rightarrow y$. Therefore if $q<h(y)$ $I\!\!D^qF(y)=0$, as
in the case of the Weierstrass function, showing that $h(y)$ is a lower
bound on the critical order. 

The procedure of finding an upper bound is technical and lengthy.
It is carried out in the Appendix below.

In this way an intricate analysis of finding out the lower bound on the
H\"older exponent has been replaced by a calculation involving few steps. This
calculation can easily be generalized for more general functions $g(x)$. 
Summarizing, the LFD enables one to calculate the local H\"older exponent even
for the case of multifractal functions. This fact, proved in theorems I and II 
is demonstrated with a concrete illustration.

\section{Conclusion}

In this paper we have introduced the notion of a local fractional derivative
using Riemann-Liouville formulation (or equivalents such as Grunwald's)
of fractional calculus. This definition was found to appear naturally
in the Taylor expansion (with a remainder) of functions and thus is
suitable for approximating scaling functions. In particular
we have pointed out a possibility of replacing the notion of a
tangent as an equivalence class of curves passing through the same point
and having the same derivative with a more general one.
This more general notion is in terms of an equivalence class of curves
passing through the same point and having the same critical order and 
the same LFD.
This generalization has the advantage 
of being applicable to non-differentiable functions also.

We have established that (for sufficiently large $\lambda$ ) the critical order of the
Weierstrass function is related to the box dimension of its graph. If  the dimension of
the graph of such a function is $1+\gamma$, the critical order is $1-\gamma$. When 
$\gamma$ approaches unity the function becomes more and more irregular and local fractional
differentiability is lost accordingly. Thus there is a direct quantitative connection between the 
dimension of the graph and the fractional differentiability property of the function.
This is one of the main conclusions of the present work.
A consequence of our result is that a classification of continuous paths 
(e.g., fractional Brownian paths) or
functions according to local fractional differentiability properties is also
a classification according to  dimensions (or H\"older exponents).

Also  the L\'evy index of a L\'evy flight on a one dimensional 
lattice is identified as
the critical order of the characteristic function of the walk. More generally,
the L\'evy index of a L\'evy  distribution is identified as
the critical order of its characteristic function at the origin.

We have argued and demonstrated that LFDs are useful for studying isolated singularities and singularities masked by the stronger singularity (not just by
regular behavior). We have further shown that the pointwise
behavior of irregular, fractal or multifractal functions can be studied
using the methods of this paper.

We hope that future study in this direction will make random irregular 
functions as well as multivariable irregular functions 
amenable to analytic treatment, which is badly needed at this
juncture. Work is in progress in this direction.

%\vspace{0.1in}
%\noindent
%{\large \bf{Acknowledgement}}
\section*{Acknowledgments}

We acknowledge helpful discussions with Dr. H. Bhate and Dr. A. Athavale.
One of the authors (KMK) is grateful to CSIR (India) for financial assistance and the other author
(ADG) is grateful to UGC (India) financial assistance during initial stages of the work.

\appendix
\section*{Appendix: Upper bound on critical order of function of Eq.~\cite{24}}
Our aim in this is appendix is to
 find an upper bound on the critical order and hence on the local
H\"older exponent for the function of equation (\ref{soln}). Our procedure will be similar to that of Jaffard \cite{24} the only difference 
being that we  take the fractional derivative instead of the wavelet transform.
We proceed as follows. The defining equation for $F(x)$ when iterated 
$N$ times gives
\begin{eqnarray}
F(x)&=&\sum_{n=0}^{N-1}\;\;\;
\sum_{i_1,\cdots,i_n=1}^2{\lambda}_{i_1}\cdots{\lambda}_{i_n}
g(S_{i_n}^{-1}\cdots S_{i_1}^{-1}(x))\nonumber\\
&+& \sum_{i_1,\cdots,i_N=1}^2{\lambda}_{i_1}\cdots{\lambda}_{i_N}
F(S_{i_N}^{-1}\cdots S_{i_1}^{-1}(x)),\;\;\;\;\;\;x\in [0,1].
\end{eqnarray}
We now consider
\begin{eqnarray}
{d^q(F(x)-F(y))\over{[d(x-y)]^q}}&=&\sum_{n=0}^{N-1}\;\;\;\sum_{i_1,\cdots,i_n=1}^2{\lambda}_{i_1}\cdots{\lambda}_{i_n}
({\mu}_{i_1}\cdots{\mu}_{i_n})^{-q} \nonumber\\
&&\;\;\;\;\;\;\;\;\;\;{d^q[g(S_{i_n}^{-1}\cdots S_{i_1}^{-1}(x))-g(S_{i_n}^{-1}\cdots S_{i_1}^{-1}(y))]
\over{[d(S_{i_n}^{-1}\cdots S_{i_1}^{-1}(x-y))]^q}}\nonumber\\
&&+\sum_{i_1,\cdots,i_N=1}^2{\lambda}_{i_1}\cdots{\lambda}_{i_N}
({\mu}_{i_1}\cdots{\mu}_{i_N})^{-q} \nonumber \\
&&\;\;\;\;\;\;\;\;\;\;{d^q[F(S_{i_N}^{-1}\cdots S_{i_1}^{-1}(x))-F(S_{i_N}^{-1}\cdots S_{i_1}^{-1}(y))]
\over{[d(S_{i_N}^{-1}\cdots S_{i_1}^{-1}(x-y))]^q}}. \label{iterated}
\end{eqnarray}
Let us denote the first term on the RHS by $A$ and the second term by $B$.
In the following $y\in (0,1)$.
Choose $N$ such that $3^{-(N+1)}<\vert x-y \vert < 3^{-N}$.
 Denote $\lambda_{n(y)}={\lambda}_{{i_1}(y)}\cdots{\lambda}_{{i_n}(y)}$.
For the values of $\mu_i$s we have chosen ${\mu}_{i_1}\cdots{\mu}_{i_n}
= 3^{-n}$.
Now since $g$ is smooth $\vert g(x)-g(y) \vert \leq C \vert x-y \vert$,
\begin{eqnarray}
{d^q[g(S_{i_n}^{-1}\cdots S_{i_1}^{-1}(x))-g(S_{i_n}^{-1}\cdots S_{i_1}^{-1}(y))]
\over{[d(S_{i_n}^{-1}\cdots S_{i_1}^{-1}(x-y))]^q}}
& \leq & C \vert {{x-y}\over{{\mu}_{i_1}\cdots{\mu}_{i_n}}} \vert^{1-q}.
\end{eqnarray}
From the way we chose $\mu_i$s and $\vert x-y \vert$ this term is bounded by
$C 3^{n(1-q)} 3^{-N(1-q)}$.
Therefore the first term in equation (\ref{iterated}) above is bounded by
\begin{eqnarray}
 A &\leq & C\sum_{n=0}^{N-1} \lambda_{n(y)} 3^{nq} 
 3^{n(1-q)} 3^{-N(1-q)}  \nonumber\\
&=& C3^{-N(1-q)}\sum_{n=0}^{N-1} \lambda_{n(y)} 3^{n} \nonumber\\
&\leq & C3^{-N(1-q)}\lambda_{(N-1)(y)} 3^{N-1} (1+{\lambda_{(N-2)(y)}\over{
3\lambda_{(N-1)(y)}}}+{\lambda_{(N-3)(y)}\over{3^2\lambda_{(N-1)(y)}}}+\cdots).
\end{eqnarray}
The quantity in the brackets is bounded. Therefore
\begin{eqnarray}
 A &\leq & C3^{Nq}\lambda_{N(y)}.  
\end{eqnarray}
Now we consider the second term on the RHS of equation (\ref{iterated}) and find a lower bound on that term.
We assume that $F(x)$ is not Lipschitz at $y$ (for otherwise the LHS of
equation (\ref{iterated}) will be zero in the limit $x\rightarrow y$ and the 
case is uninteresting).
Therefore there exists a sequence of points $x_n$ approaching $y$ such that
\begin{eqnarray}
\vert F(x_n)-F(y) \vert \geq c \vert x_n - y \vert. \label{nonsmooth}
\end{eqnarray} 
That the function is not Lipschitz at $y$ implies that it not Lipschitz at
$S_{i_N}^{-1}\cdots S_{i_1}^{-1}(y)$. Therefore we get (one may recall
that only one of the $2^n$ terms in the summation 
$\sum_{i_1,\cdots,i_N=1}^2$ is non-zero)
\begin{eqnarray}
B&\geq & c\lambda_{N(y)} 3^{Nq} 
 3^{N(1-q)} 3^{-N(1-q)} \nonumber\\
&\geq & c \lambda_{N(y)} 3^{Nq}.
\end{eqnarray}
Therefore there exists a sequence $x_n$ such that
\begin{eqnarray}
\vert {d^q(F(x_n)-F(y))\over{[d(x_n-y)]^q}}-c \lambda_{N(y)} 3^{Nq} \vert
\leq  C \lambda_{N(y)} 3^{Nq}.
\end{eqnarray}
Since (\ref{nonsmooth}) is valid for every $c$ for large enough $n$,
$${d^q(F(x_n)-F(y))\over{[d(x_n-y)]^q}}\geq C' \lambda_{N(y)} 3^{Nq},$$
where $C'=c-C$.
Now from the expression of $h(y)$ it is clear that when $q>h(y)$ 
LFD does not exist and the critical order is bounded from above by $h(y)$.

%\newpage
%\section*{References}

\end{document}